\documentclass[a4paper,reqno,12pt]{amsart}

\usepackage[utf8]{inputenc}
\usepackage[T1]{fontenc}
\usepackage{lmodern}
\usepackage[UKenglish]{babel}
\usepackage[UKenglish]{isodate}
\usepackage{amsmath, amsfonts, amssymb, amsthm, mathtools, cancel, bm}
\usepackage{enumerate}
\usepackage{stmaryrd}
\usepackage{empheq}
\usepackage{geometry} 
\usepackage{mathrsfs} 
\usepackage{hyperref} 
\hypersetup{colorlinks = true}
\hypersetup{linktocpage} 
\usepackage[capitalise]{cleveref} 
\usepackage{dsfont}
\usepackage{calligra}
\usepackage{braket}

\theoremstyle{plain}
\newtheorem{thm}{Theorem}[section]

\theoremstyle{definition}

\theoremstyle{remark} 
\newtheorem{rmkx}[thm]{Remark}

\usepackage{dsfont}
\usepackage{geometry}
\makeatletter
\newtheoremstyle{plain}
  {8pt plus 2pt minus 4pt}
  {8pt plus 2pt minus 4pt}
  {\itshape}
  {}
  {\bfseries}
  {.}
  {5pt plus 1pt minus 1pt}
  {}
\makeatother

\theoremstyle{plain}
\newtheorem{theorem}{Theorem}[section]
\newtheorem{remark}{Remark}[section]

\def \R {{\mathbb {R}}}

\def \C {{\mathbb C}}

\def \Z {{\mathbb Z}}

\newcommand{\diff}{\mathop{}\mathopen{}\mathrm{d}}
\DeclareMathOperator{\tr}{tr}
\newcommand{\Hdiv}{\dot{H}^1_{\mathrm{div}}\left(\R^3\right)} 
\newcommand{\setst}[2]{\left\{#1\mathrel{}\middle|\mathrel{}#2\right\}}
\newcommand{\norm}[1]{\left\lVert#1\right\rVert}
\DeclareMathOperator{\curl}{curl}
\newcommand{\abs}[1]{\left\lvert#1\right\rvert}
\newcommand{\bigO}[1]{O\mathopen{}\left(#1\right)}
\newcommand{\schatten}[1]{\mathfrak{S}_{#1}}

\title[The Electromagnetic Dirac Field at Thermal Equilibrium]{Recent Advances on a Model \\ for the Electromagnetic Dirac Field \\ at Thermal Equilibrium}
\author[Umberto Morellini]{Umberto Morellini}
\address{Università di Pisa, Dipartimento di Matematica, Largo Bruno Pontecorvo 5, 56127 Pisa, Italy}
\email{\href{mailto:umberto.morellini@dm.unipi.it}{\nolinkurl{umberto.morellini@dm.unipi.it}}}
\date{23 July 2026}

\begin{document}

\begin{abstract}
   We review recent progress on a model for the relativistic electron-positron field interacting with a classical electromagnetic field at positive temperature. The ultimate goal is to advance our comprehension of how thermal backgrounds affect polarisation mechanisms of this quantum system.
\end{abstract}

\maketitle

\tableofcontents

\addtocontents{toc}{\protect\setcounter{tocdepth}{1}}
\section{Introduction}
Quantum Electrodynamics (QED) is a quantum field theory which combines quantum mechanics with special relativity and aims at describing the interactions between matter and light at a scale where quantum effects are dominant. QED is formulated as a perturbative theory which makes it unsatisfactory from a mathematical point of view. For this reason, several approximations have been addressed in the past decades.\par
QED in a lattice was investigated in \cite{OstSei-1978-AP, Sei-1982-book}, while linear and non-linear Dirac operator models for finitely many particles in a classical electromagnetic field and an `inert' vacuum were studied in \cite{DolEstLos-2007-AHP, EstGeoSer-1996-CVPDE, EstSer-1999-CMP, FlaSimTaf-1997-MAMS, Geo-1991-IUMJ, HubSie-2007-ARMA} (see also the references in \cite{EstLewSer-2008-BAMS}). In addition, the quantum vacuum and the process of pair creation in a non-interacting setting — meaning without light — were analysed in \cite{KlaSch-1977-HPAa, KlaSch-1977-HPAb, Nen-1987-CMP, NenSch-1978-HPA, PicDur-2008-CMP}.\par
Recently, an alternative approach based on a Hartree-Fock-type model was developed in a series of works \cite{GraHaiLewSer-2013-ARMA, GraLewSer-2009-CMP, GraLewSer-2011-CMP, GraLewSer-2018-JMPA, HaiLewSer-2005-CMP, HaiLewSer-2005-JPA, HaiLewSer-2009-ARMA, HaiLewSerSol-2007-JPA, HaiLewSol-2007-CPAM}, originating from the foundational papers of Chaix and Iracane \cite{ChaIra-1989-JPB, ChaIraLio-1989-JPB} and building on \cite{BacBarHelSie-1999-CMP, HaiSie-2003-CMP}. This framework models relativistic particles in a fluctuating vacuum — both described by the Dirac operator — interacting with a classical electromagnetic field. While the lack of light quantisation omits certain physical effects, these works provide the first mathematical results for a quantum vacuum interacting with light. Furthermore, they demonstrate that the associated model can be constructed in a fully non-perturbative, though simplified, setting.\par
It is worth noticing that Hartree-Fock-type QED models have additionally found application in investigating relativistic molecular dynamics \cite{HaiLewSpa-2005-LMP, Mor-2025-JDE}, whereas a two-dimensional counterpart has been employed to study graphene \cite{BorMor-2025-JPA, HaiLewSpa-2012-JMP}. Furthermore, a one-dimensional variant has been recently numerically explored as an initial step towards a concrete implementation of an effective QED theory for atoms and molecules \cite{AudMorLevTou-2025-JPA}.\par
As described below, in various situations it is interesting to consider the same kind of problems in a thermal background. A first result in this direction is represented by \cite{HaiLewSei-2008-RMP}, where the purely electrostatic case is addressed. This article aims at reviewing some recent advances which have been done in the study of the relativistic electron-positron field at positive temperature, first in the purely magnetic case \cite{Mor-2026-AHP} and then in the complete electromagnetic setting \cite{BorMor-2026-CAOT}.

\subsection{The Dirac Operator}
While considering relativistic effects in quantum mechanics, the Dirac operator naturally appears in place of the Schrödinger operator. We work in a system of units such that the speed of light and the reduced Planck constant are both set equal to one, $c=\hbar=1$. Given an electrostatic potential $V:\R^3\to \R$ and a magnetic potential $A: \R^3\to\R^3$, we define the Dirac operator with mass $m$, elementary charge $e$ and electromagnetic $4$--potential $\boldsymbol{A}=\left(V,A\right)$ as
\begin{equation*}
    D_m^{e\boldsymbol{A}}=\boldsymbol{\alpha}\cdot\left(-i\nabla-eA\right)+eV+m\beta\,,
\end{equation*}
acting on $L^2\left(\R^3,\C^4\right)$. The four Dirac matrices $\boldsymbol{\alpha}=\left(\alpha_1,\alpha_2,\alpha_3\right)$ and $\beta$ are given by
\[
\alpha_k=\begin{bmatrix}
0 & \sigma_k\\
\sigma_k & 0
\end{bmatrix}\,,\quad
\beta=\begin{bmatrix}
I_2 & 0\\
0 & -I_2
\end{bmatrix}\,,
\]
with Pauli matrices $\sigma_1$, $\sigma_2$ and $\sigma_3$ defined as
\[
\sigma_1=\begin{bmatrix}
0 & 1\\
1 & 0
\end{bmatrix}\,,\quad
\sigma_2=\begin{bmatrix}
0 & -i\\
i & 0
\end{bmatrix}\,,\quad
\sigma_3=\begin{bmatrix}
1 & 0\\
0 & -1
\end{bmatrix}\,.
\]
The \emph{free} Dirac operator $D^0_m$, \emph{i.e.} with $\boldsymbol{A}\equiv 0$, is self-adjoint on the Sobolev space $H^1\left(\R^3,\C^4\right)$, and its spectrum is purely absolutely continuous and given by
\begin{equation*}
\sigma(D^0_m)=(-\infty,-m]\cup[m,+\infty)\,.
\end{equation*}
We refer the reader to \cite{Tha-1992-book} for further details.\par
Since in quantum mechanics it is always assumed that the most stable state of the system is the one with the lowest energy, the fact that the Dirac operator is unbounded from below — meaning that the energy can be arbitrarily negative — is highly surprising. From a mathematical point of view, this makes a huge difference as compared to non-relativistic models based on the Laplace operator $-\Delta$. Since Dirac did not want to renounce the physical picture where states with lower energy are more stable, he postulated that the vacuum is filled with infinitely many virtual particles occupying the negative energy states (\emph{Dirac sea}) \cite{Dir-1930-PRSL, Dir-1934-MPCPS, Dir-1934-SR}. As a consequence, a real free electron cannot be in a negative state due to the Pauli exclusion principle. Moreover, Dirac also conjectured the existence of `holes' in the Dirac sea interpreted as positrons, having a positive charge and a positive energy, which were later experimentally discovered by Anderson \cite{And-1933-PR}. Furthermore, the theory predicts the phenomenon of vacuum polarisation: while the free vacuum is electrically neutral, in presence of an external electrostatic field the virtual electrons respond to the applied field and the vacuum acquires a non-constant density of charge. In addition, the polarised vacuum itself modifies the electrostatic field and the virtual electrons react to the corrected field. For a detailed review of the challenges posed by the negative spectrum of the Dirac operator, see \cite{EstLewSer-2008-BAMS}. 

\subsection{Physical Context at Positive Temperature}
The non-linear polarisable nature of the vacuum in quantum field theory is a well-established fact since a long time \cite{Dir-1934-MPCPS, EulKoc-1935-DN, Wei-1936-MFM}. In this context, one replaces the classical action with an effective one accounting for quantum corrections, still guaranteeing the validity of the principle of least action. In QED, a well-known way to implement this strategy is by integrating out fermions in the path integral formulation (see for instance \cite[Ch.~33]{Sch-2013-book}). As a result, a functional of a classical electromagnetic field treated as an external one is obtained.\par
The study of positive-temperature effects in quantum field theory attracted a considerable interest in the Physics literature in the past decades.
Indeed, in several physical examples, the fields exist in a thermal bath or in some non-equilibrium background which differs from the vacuum. Several authors \cite{DolJac-1974-PRD, KirLin-1972-PLB, Wei-1974-PRD} considered systems of elementary particles described by a quantum field theory in a positive-temperature background. They found that while specific symmetries are spontaneously broken at zero temperature (such as those concerning weak interactions), they may be restored at sufficiently high temperatures. Moreover, the critical temperature characterising such a phase transition has been calculated. To do so, one needs to know the Feynman rules for a field theory at finite temperature. For a non-gauge theory, they can be derived using well-known methods \cite{FetWal-1971-book}. However, for a gauge theory such as QED, various problems arise and a more powerful technique is needed  \cite{Ber-1974-PRD}. We mention that, in any case, the Feynman rules at finite temperature $T$ are strictly related to the Kubo-Martin-Schwinger (KMS) condition (see also Remark~\ref{stat:KMS-condition}) and concretely consist of the zero-temperature rules with the following \textit{formal} replacements:
\begin{align}\notag
    \int\frac{\diff^4 p}{\left(2\pi\right)^4}&=\frac{i}{\beta}\sum_n\int\frac{\diff^3 p}{\left(2\pi\right)^3}\,,\\\label{eq:T-Feynman-rules}
    p_0&=i\omega_n\,,\\\notag
    \left(2\pi\right)^4\delta^4\left(p_1+p_2+\ldots\right)&=\frac{\beta}{i}\left(2\pi\right)^3\delta\left(\omega_{n_1}+\omega_{n_2}+\ldots\right)\delta^3\left(\Vec{p_1}+\Vec{p_2}+\ldots\right)\,,
\end{align}
where $\beta=1/T$ and
\[
\omega_n=
\begin{cases}
    \frac{2n\pi i}{\beta}\quad\text{for bosons}\,,\\
    \frac{\left(2n+1\right)\pi i}{\beta}\quad\text{for fermions}\,,
\end{cases}
\]
depending on the statistics of the particles involved.\par
Thermal effects in quantum electrodynamics have been first studied in \cite{Dit-1979-PRD}, where the author calculates the one-loop effective potential at finite temperature for scalar and spinor QED in a constant magnetic field — effectively extending the Euler-Heisenberg Lagrangian to positive temperatures. Thermodynamically, this potential represents the vacuum energy contribution to the total free energy. Later, this thermal effective action has been analysed under more general assumptions to additionally include a chemical potential \cite{Dit-1979-PRD, ElmPerSka-1993-PRL, ElmSka-1995-PLB, Sch-1951-PR}. For a complete review about QED at finite temperature, see \cite[Chapter~3.5]{DitGie-2000-book}.

\section{The Model}
In this section, we recall the definition of the Pauli-Villars-regularised free energy of the Dirac field ad thermal equilibrium, which has been rigorously studied in \cite{BorMor-2026-CAOT, Mor-2026-AHP} where a more detailed derivation can also be found.\par
The Dirac field free energy $\mathcal{F}_\mathrm{eq}\left(e\boldsymbol{A},\beta\right)$ at thermal equilibrium is \emph{formally} given by
\begin{equation}\label{eq:vacuum-free-energy}
    \mathcal{F}_\mathrm{eq}\left(e\boldsymbol{A},\beta\right)\coloneqq-\frac{1}{\beta}\tr\left[\log\left(2\cosh\frac{\beta D^{e\boldsymbol{A}}_m}{2}\right)\right]\,.
\end{equation}
Because of the unboundedness of the electromagnetic Dirac operator $D^{e\boldsymbol{A}}_m$, the trace in \eqref{eq:vacuum-free-energy} is of course infinite, except if our model is settled in a box with an ultraviolet cut-off \cite{HaiLewSol-2007-CPAM}. In order to give a proper mathematical meaning to $\mathcal{F}_\mathrm{eq}\left(e\boldsymbol{A},\beta\right)$, we can first subtract the (infinite) free energy of the free Dirac field corresponding to $\boldsymbol{A}\equiv 0$ and consider the \emph{relative} free energy given by
\begin{equation}\label{eq:relative-free-energy}
    \mathcal{F}_\mathrm{rel}\left(e\boldsymbol{A},\beta\right)\coloneqq \frac{1}{\beta}\tr\left[\log\left(2\cosh\frac{\beta D^{0}_m}{2}\right)-\log\left(2\cosh\frac{\beta D^{e\boldsymbol{A}}_m}{2}\right)\right]\,.
\end{equation}
From a variational viewpoint, one can expect that removing an infinite constant does not alter the problem under review.\par
However, the functional \eqref{eq:relative-free-energy} is not well-defined yet, due to ultraviolet (UV) divergences which represent a well-known issue in QED and related models. Indeed, the operator $\log\left(2\cosh\left(\beta D^{0}_m/2\right)\right)-\log\left(2\cosh\left(\beta D^{e\boldsymbol{A}}_m/2\right)\right)$ is not trace-class as long as $\boldsymbol{A}\neq 0$, and this can be formally seen by expanding the trace in a power series of $e\boldsymbol{A}$: the first-order term in the expansion vanishes, while the second-order term is infinite due to UV logarithmic divergences (see Theorem~\ref{stat:properties-FPV} and Section~\ref{sec:main-thm-2}). Since the higher-order terms cannot prevent this issue, an UV cut-off is needed.\par
In this regard, the choice of the regularisation is extremely important. As shown in Section~\ref{sec:main-thm-2}, thanks to the gauge invariance, several terms in the expansion of $\eqref{eq:relative-free-energy}$ as a power series of $e\boldsymbol{A}$ vanish. This is a further reason to preserve the gauge symmetry, in addition to its well-known physical meaning. In \cite{GraLewSer-2009-CMP}, the authors studied two approaches to handle UV divergences in the purely electrostatic case by imposing a cut-off. However, these methods are not applicable here, since they do not preserve gauge symmetry.\par
Thus, we are going to use the regularisation technique introduced by Pauli and Villars \cite{PauVil-1949-RMP} in $1949$. Of course, this is not the only possible choice (see for instance \cite{Lei-1975-RMP} for an alternative approach). The mentioned Pauli-Villars (PV) method consists of introducing two fictitious particles into the model with very high masses $m_1,m_2$, essentially playing the role of ultraviolet cut-offs. Indeed, note that the mass has the dimension of a momentum since in our system of units $\hbar=c=1$. The sole purpose of these additional particles is to regularise the model at high energies, and they have no direct physical interpretation. Moreover, being very massive, they do not affect the low-energy regime. Therefore, we consider the so-called Pauli-Villars-regularised free energy of the relativistic electron-positron field at thermal equilibrium defined as
\begin{equation}\label{eq:PV-free-energy}
    \widetilde{\mathcal{F}}_\mathrm{PV}\left(e\boldsymbol{A},\beta\right)\coloneqq\frac{1}{\beta}\tr\left[\sum_{j=0}^2 c_j\left(\log\left(2\cosh\frac{\beta D^{0}_{m_j}}{2}\right)-\log\left(2\cosh\frac{\beta D^{e\boldsymbol{A}}_{m_j}}{2}\right)\right)\right]\,.
\end{equation}
Here, the coefficient $c_0$ and the mass $m_0$ are respectively equal to $1$ and $m$. The corresponding term is exactly the relative free energy in \eqref{eq:relative-free-energy}. The ultraviolet divergences are removed if the coefficients $c_1,c_2$ fulfil the PV conditions
\begin{equation}\label{eq:PV-conditions}
    \sum_{j=0}^2 c_j=\sum_{j=0}^2 c_j m_j^2=0\,,
\end{equation}
which are equivalent to
\begin{equation*}
    c_1=\frac{m_0^2-m_2^2}{m_2^2-m_1^2}\quad\text{and}\quad c_2=\frac{m_1^2-m_0^2}{m_2^2-m_1^2}\,.
\end{equation*}
Furthermore, we shall always assume that $m_0<m_1<m_2$, which implies that $c_1<0$ and $c_2>0$. The role of the constraint \eqref{eq:PV-conditions} is to remove the worst \emph{linear} ultraviolet divergences. Indeed, the PV regularisation does not avoid yet a logarithmic divergence when $m_1,m_2$ go to infinity. This can be understood by defining the averaged UV cut-off $\Lambda$ as
\begin{equation}\label{eq:averaged-UV-cutoff}
    \log\left(\Lambda^2\right)=-\sum_{j=0}^2 c_j\log\left(m_j^2\right)\,.
\end{equation}
Observe that, once the value of $\Lambda$ is fixed, the values of $m_1,m_2$ are not uniquely determined. Concretely, one usually chooses the masses as functions of the cut-off $\Lambda$, so that the coefficients $c_1,c_2$ stay bounded when $\Lambda\xrightarrow{}\infty$. The persistent logarithmic divergence in the averaged cut-off parameter $\Lambda$ can be explicitly seen in the second-order term in the expansion (see Equation~\eqref{eq:log-divergence}).  \par
Now, it is convenient to rewrite the Pauli-Villars-regularised free energy functional \eqref{eq:PV-free-energy} in an integral form. For this purpose, we differentiate it with respect to $\beta$
\[
\frac{\partial}{\partial\beta}\left(\beta \widetilde{\mathcal{F}}_\mathrm{PV}\left(e\boldsymbol{A},\beta\right)\right)=\frac{1}{2}\tr\left[\sum_{j=0}^2 c_j \left(D^{0}_{m_j} \tanh\frac{\beta D^{0}_{m_j}}{2}-D^{e\boldsymbol{A}}_{m_j} \tanh\frac{\beta D^{e\boldsymbol{A}}_{m_j}}{2}\right)\right]\,,
\]
which implies
\begin{equation}\label{eq:integral-free-energy}
    \widetilde{\mathcal{F}}_\mathrm{PV}\left(e\boldsymbol{A},\beta\right)=\frac{1}{2\beta}\int_0^\beta \tr\left[\sum_{j=0}^2 c_j \left(D^{0}_{m_j} \tanh\frac{b D^{0}_{m_j}}{2}-D^{e\boldsymbol{A}}_{m_j} \tanh\frac{b D^{e\boldsymbol{A}}_{m_j}}{2}\right)\right]\diff b\,.
\end{equation}
However, we still need to modify the free energy functional: following the study of the electromagnetic case at zero temperature \cite{GraHaiLewSer-2013-ARMA}, we will rather work with
\begin{equation}\label{eq:boxed-PV-free-energy}
\boxed{
 \mathcal{F}_\mathrm{PV}\left(e\boldsymbol{A},\beta\right)=\frac{1}{2\beta}\int_0^\beta \tr\left(\tr_{\C^4}\left[\sum_{j=0}^2 c_j \left(D^{0}_{m_j} \tanh\frac{b D^{0}_{m_j}}{2}-D^{e\boldsymbol{A}}_{m_j} \tanh\frac{b D^{e\boldsymbol{A}}_{m_j}}{2}\right)\right]\right)\diff b\,.}
\end{equation}
Notice that the current integrand includes an extra $\C^4$--trace compared with Equation~\eqref{eq:integral-free-energy}. According to Theorem~\ref{stat:definition-FPV}, this modification guarantees that the resulting operator remains trace-class in the presence of the electrostatic potential. While one could also ensure this by increasing the number of fictitious particles in the Pauli-Villars regularisation and imposing additional mass constraints (see the comment after \cite[Theorem~2.1]{GraHaiLewSer-2013-ARMA} for further details), we prefer to adapt the functional using two auxiliary masses, as typically done in the Physics literature.

\section{Well-Posedness and Properties of the Free Energy Functional}
The first result shows that the Pauli-Villars-regularised free energy functional $\boldsymbol{A} \mapsto \mathcal{F}_\mathrm{PV}\left(e\boldsymbol{A},\beta\right)$ in \eqref{eq:boxed-PV-free-energy} is well-defined, albeit under provisional non-gauge-invariant assumptions. In other words, the hypotheses are imposed directly on the vector potential rather than on the physical field.
\begin{theorem}[Definition of $\mathcal{F}_\mathrm{PV}$ in energy space]\label{stat:definition-FPV}
    Assume that the constants $c_j$ and $m_j$, $j=0,1,2$, satisfy
    \[
    c_0=1\,,\quad m_2>m_1>m_0>0\,,
    \]
    and \eqref{eq:PV-conditions}.
    Let
    \begin{equation*}
    T_{\boldsymbol{A}}\left(\beta\right)=\frac{1}{2}\sum^2_{j=0}c_j\left( D^0_{m_j}\tanh{\left(\frac{\beta D^0_{m_j}}{2} \right)}- D^{ \boldsymbol{A}}_{m_j}\tanh{\left(\frac{\beta D^{\boldsymbol{A}}_{m_j}}{2}\right)} \right)\,,
    \end{equation*}
    with $\boldsymbol{A}=\left(V,A\right)$. Then
    \begin{itemize}
    \item[i)] For every $\boldsymbol{A}\in L^1\left(\R^3,\R^4\right)\cap H^1\left(\R^3,\R^4\right)$, the operator $\tr_{\C^4}T_{\boldsymbol{A}}\left(\beta\right)$ is trace-class on $L^2\left(\R^3,\C\right)$. In particular, the functional $\mathcal{F}_\mathrm{PV}\left(e\boldsymbol{A},\beta\right)$ is well-defined as
    \begin{equation}\label{eq:FPV}
        \boxed{
        \mathcal{F}_\mathrm{PV}\left(e\boldsymbol{A},\beta\right)=\frac{1}{\beta}\int^\beta_0 \tr\left(\tr_{\C^4}T_{\boldsymbol{A}}\left(b\right) \right)\diff b\,.}
    \end{equation}
    \item[ii)] If $V\equiv 0$ and $A\in L^1\left(\R^3,\R^4\right)\cap H^1\left(\R^3,\R^4\right)$, the operator $T_{\left(0,A\right)}\left(\beta\right)$ is trace-class on $L^2\left(\R^3,\C^4\right)$, hence
    \begin{equation*}
        \mathcal{F}_\mathrm{PV}\left(0,eA,\beta\right)=\frac{1}{\beta}\int^\beta_0 \tr\left(T_{\left(0,A\right)}\left(b\right) \right)\diff b\,.
    \end{equation*}
    \end{itemize}
\end{theorem}
\begin{remark}
An analogue of Theorem~\ref{stat:definition-FPV} has been previously established at zero temperature in \cite[Theorem~2.1]{GraHaiLewSer-2013-ARMA}. There, the authors observe that, while the $\C^4$--trace is required to eliminate non-trace-class terms in the presence of a generic electrostatic potential, such a regularisation `trick' becomes unnecessary when $V \equiv 0$. Our result proves that this feature persists at finite temperature, confirming that the reduction to two auxiliary masses via the preliminary $\C^4$--trace remains a mathematically sound and convenient strategy also in the thermal regime.
\end{remark}\par
Let us now introduce the energy space
\begin{multline*}
    \Hdiv\coloneqq\big\{\boldsymbol{A}=\left(V,A\right)\in L^6\left(\R^3,\R^4\right)\big| \mathrm{div}A=0\\
    \text{ and }\boldsymbol{F}=\left(-\nabla V,\mathrm{curl}A\right)\in L^2\left(\R^3,\R^6\right)\big\}\,,
\end{multline*}
endowed with the norm
\begin{equation*}
    \|\boldsymbol{A}\|^2_{\Hdiv}=\|\nabla V\|^2_{L^2\left(\R^3\right)}+\|\mathrm{curl}A\|^2_{L^2\left(\R^3\right)}=\|\boldsymbol{F}\|^2_{L^2\left(\R^3\right)}\,.
\end{equation*}
Let us stress that $\boldsymbol{F}=(E,B)$ corresponds to the electromagnetic field associated to the $4$--potential $\boldsymbol{A}$. A further investigation into the properties of $\mathcal{F}_\mathrm{PV}$ requires the assumption of an integrable electric field, $E=-\nabla V\in L^1\left(\R^3,\R^3\right)$, which is clearly gauge-invariant.
\begin{theorem}[Properties of $\mathcal{F}_\mathrm{PV}$]\label{stat:properties-FPV}
    Let the constants $c_j$ and $m_j$, $j=0,1,2$, satisfy the same assumptions as in Theorem~\ref{stat:definition-FPV}.
    \begin{itemize}
        \item[i)] Let $\boldsymbol{A}\in L^1\left(\R^3,\R^4\right)\cap\Hdiv$ and $E=-\nabla V\in L^1\left(\R^3,\R^3\right)$. Then, we have
        \begin{equation}\label{eq:FPV-decomposition}
            \mathcal{F}_\mathrm{PV}\left(e\boldsymbol{A},\beta\right)=\mathcal F_2\left(\boldsymbol{F},\beta\right)+\mathcal{R}\left(\boldsymbol{A},\beta\right)\,,
        \end{equation}
        where $\boldsymbol{F}=\left(E,B\right)$, with $B=\curl A$. The functional $\mathcal{R}$ is continuous on $\Hdiv$ and there exists a constant $\kappa=\kappa\left(\beta\right)$ such that 
        \begin{equation}\label{eq:R-bound}
            \vert \mathcal{R}\left(\boldsymbol{A},\beta\right)\vert \leq \kappa\left(\left(\sum_{j=0}^2\frac{\abs{c_j}}{m_j}\right)\norm{\boldsymbol{F}}^4_{L^2\left(\R^3\right)}+\left(\sum_{j=0}^2\frac{\abs{c_j}}{m_j^2}\right)\norm{\boldsymbol{F}}^6_{L^2\left(\R^3\right)}\right)\,.
        \end{equation}
        \item[ii)] The functional $\mathcal{F}_2$ is the bounded quadratic form on $L^1\left(\R^3,\R^4\right) \cap  L^2\left(\R^3,\R^4\right)$ given by
        \begin{empheq}[box=\fbox]{align} \label{eq:F2}
        \begin{split}
            \mathcal{F}_2\left(\boldsymbol{F},\beta\right)=\frac{1}{8\pi}\int_{\R^3}\left(M^0\left(k\right)+M^T\left(k,\beta\right)\right)&\left(\vert \widehat B\left(k\right)\vert^2-\vert \widehat E\left(k\right)\vert^2\right)\, \diff k\\
            &\hspace{-1cm}+ \frac{1}{\beta}\int^\beta_0\int_{\R^3}\frac{\Gamma\left(k,b\right)}{\vert k\vert^2}\vert \widehat E\left(k\right)\vert^2\,\diff k\,\diff b\,,
            \end{split}
        \end{empheq}
        where
        \begin{equation}\label{eq:Mzero}
        M^0\left(k\right)=-\frac{2}{\pi}\int^1_0\sum^2_{j=0}c_j\log\left(m^2_j+u\left(1-u\right)k^2\right)\, u\left(1-u\right)\,\diff u\,,
        \end{equation}
        \begin{equation}\label{eq:MT}
            \begin{split}
                &M^T\left(k,\beta\right)=-\frac{8}{\pi}\int^1_0 \diff u\, u\left(1-u\right)\int^\infty_0\,\diff t\,\sum^2_{j=0}c_j \left(\frac{1}{1+e^{-X_j\left(\beta,u,k\right)\cosh t} }\right. \\
                &\quad\left.+ \frac{1}{X_j\left(\beta,u,k\right)\cosh t}\left(2\log2-\log\left(\left(1+e^{X_j\left(\beta,u,k\right)\cosh t}\right) \left(1+e^{-X_j\left(\beta,u,k\right)\cosh t}\right) \right) \right)\right)\,,
            \end{split}
        \end{equation}
        with $X_j\left(\beta,u,k\right)=\beta\sqrt{m_j^2+u\left(1-u\right)k^2}$ for $j=0,1,2$, and $\Gamma \in C^0\left(\R^3\times\left(0,+\infty\right)\right)$ is an explicit function of $\left(k,\beta\right)$ (see \eqref{eq:Gamma}), such that, for any $\beta\in\left(0,+\infty\right)$,
        \begin{equation*}
            \frac{\Gamma\left(\cdot,\beta\right)}{\vert \cdot\vert^2}\in L^1_{\mathrm{loc}}\left(\R^3\right)\,,\quad \mbox{and}\quad \frac{\Gamma\left(\cdot,\beta\right)}{\vert \cdot\vert^2}\in L^\infty\left(\{\vert k\vert\geq \delta\}\right)\,,\quad\forall \delta>0\,.
        \end{equation*}
        \item[iii)] The function $M^0$ is well-defined and positive on $\R^3$, and satisfies
        \begin{equation}\label{eq:log-divergence}
            0<M^0\left(k\right)\leq M^0\left(0\right)=\frac{2\log\Lambda}{3\pi}\,,
        \end{equation}
        where $\Lambda$ is defined by \eqref{eq:averaged-UV-cutoff}. Moreover, $M^T$ is well-defined, bounded for any fixed $\beta\in\left(0,+\infty\right)$, and such that
        \begin{equation*}
            M^0\left(k\right)+M^T\left(k,\beta\right)\geq 0\,.
        \end{equation*}
    \end{itemize}
\end{theorem}
In \eqref{eq:FPV-decomposition}, we demonstrate that $\mathcal{F}_\mathrm{PV}$ in \eqref{eq:FPV} can be rewritten as the sum of a principal term involving the fields and a remainder. Although the latter depends on the $4$--potential, it can be suitably estimated in terms of the electromagnetic field (see \eqref{eq:R-bound}).
\begin{remark}\label{rmk:quadratic-term}
    The quadratic term $\mathcal{F}_2$, in the form given by \eqref{eq:F2}, captures the linear response of the Dirac field to the external electromagnetic field at thermal equilibrium. The Fourier multiplier $M^0\left(k\right)$ represents the zero-temperature contribution and is well-known in the Physics literature \cite[Eq.~(5.39)]{GreRei-2009-book}. In \cite{GraHaiLewSer-2013-ARMA}, the authors prove that
    \begin{equation*}
        \lim_{\Lambda\to\infty}\left(\frac{2 \log\Lambda}{3\pi}- M^0(k) \right)=U(k):=\frac{\vert k\vert^2}{4\pi}\int^1_0\frac{z^2-z^4/3}{1+\vert k\vert^2(1-z^2)/4}\,\diff z\,,
    \end{equation*}
    where the function $U$ was first computed by Serber \cite{Ser-1935-PR} and Uehling \cite{Ueh-1935-PR}. The same function appears in previous mathematical works dealing with purely electrostatic potentials \cite{GraLewSer-2011-CMP, HaiLewSer-2005-JPA, HaiSie-2003-CMP}. On the other hand, the thermal contributions to the linear response of the Dirac field are represented by the Fourier multipliers $M^T\left(k,\beta\right)$ and $\Gamma\left(k,\beta\right)$. While the former is computed in \cite{Mor-2026-AHP} for the purely magnetic case, the latter is derived in \cite{BorMor-2026-CAOT}. In contrast to the other two terms, this contribution features a potential singularity at zero momentum (see item $ii)$ in Theorem~\ref{stat:properties-FPV}). Rather than a simple polarisation phenomenon, this singularity points toward a full screening effect typical of Plasma Physics \emph{(Debye screening)}, where the external charge is completely shielded by the field response. This phenomenon has been rigorously established in \cite{HaiLewSei-2008-RMP} in the purely electrostatic case, where the authors show that the total potential becomes short-range. This behaviour is expected to hold also when an additional magnetic field is introduced, and verifying this hypothesis constitutes the main motivation behind our research. 
\end{remark}
Finally, one can prove that the PV-regularised vacuum free energy functional $\mathcal{F}_\mathrm{PV}$ in \eqref{eq:boxed-PV-free-energy} is well-defined under fully gauge-invariant assumptions.
\begin{theorem}[Unique extension of $\mathcal{F}_\mathrm{PV}$]\label{stat:extension-FPV}
    Let the constants $c_j$ and $m_j$, $j=0,1,2$, satisfy the same assumptions as in Theorem~\ref{stat:definition-FPV}. The functional $\mathcal{F}_\mathrm{PV}$ can be uniquely extended to a continuous mapping on
    \begin{equation*}
        X\coloneqq\setst{\boldsymbol{A}=\left(V,A\right)\in \Hdiv}{E=-\nabla V\in L^1\left(\R^3,\R^3\right)}\,.
    \end{equation*}
\end{theorem}

\section{Sketch of the Proofs}
In this section, we outline the main steps of the proofs of our main theorems. For the sake of brevity, we restrict our discussion here to an overview of the strategy, as the complete arguments involve highly technical and lengthy calculations. We refer the interested reader to \cite{BorMor-2026-CAOT, Mor-2026-AHP} for the fully detailed proofs.\par
The proofs proposed in \cite{BorMor-2026-CAOT, Mor-2026-AHP} allow most of the arguments in \cite{GraHaiLewSer-2013-ARMA, GraLewSer-2018-JMPA} to be adapted, by noticing a similarity between the integral representation of $x\mapsto|x|$, which is used in the zero-temperature case to rewrite the kinetic energy functional (see \cite[Equations~(3.2)~and~(3.3)]{GraHaiLewSer-2013-ARMA}), and the power series expansion of $x\mapsto x\tanh x$ appearing in the free energy functional \eqref{eq:boxed-PV-free-energy}. To highlight this key insight, our starting point is the formula
\begin{align}
    \notag x\tanh{x}&=\sum_{\ell \in\Z}\frac{4x^2}{\left(2\ell-1\right)^2\pi^2+4x^2}\\
    \label{eq:xtanhx}&=\frac{1}{2}\sum_{\ell\in\Z}\left(2-\frac{i\left(2\ell-1\right)\pi}{2x+i\left(2\ell-1\right)\pi}+\frac{i\left(2\ell-1\right)\pi}{2x-i\left(2\ell-1\right)\pi}\right)\,.
\end{align}
Given $S$ a self-adjoint operator on $L^2\left(\R^3,\R^4\right)$ with domain $\mathrm{Dom}\left(S\right)$, the functional calculus, together with \eqref{eq:xtanhx}, yields
\begin{equation*}
    S\tanh{S}=\frac{1}{2}\sum_{\ell\in\Z}\left(2-\frac{i\left(2\ell-1\right)\pi}{2S+i\left(2\ell-1\right)\pi}+\frac{i\left(2\ell-1\right)\pi}{2S-i\left(2\ell-1\right)\pi}\right).
\end{equation*}
We notice that this series is convergent as an operator from $\mathrm{Dom}\left(S^2\right)$ to the ambient Hilbert space, since
\begin{equation*}
    \norm{\frac{4S^2}{4S^2+\left(2\ell-1\right)^2\pi^2}}_{D\left(S^2\right)\xrightarrow{}L^2\left(\R^3,\R^4\right)}\leq\min\left\{1,\left(\left(2\ell-1\right)\pi\right)^{-2}\norm{4S^2}_{D\left(S^2\right)\xrightarrow{}L^2\left(\R^3,\R^4\right)}\right\}\,.
\end{equation*}
Therefore, setting
\begin{equation*}
    \omega\left(\ell,\beta\right)=\frac{\left(2\ell-1\right)\pi}{\beta}\,,\qquad \ell\in\Z\,,
\end{equation*}
we can write
\begin{multline}\label{eq:TA-resolvent-difference}
    T_{\boldsymbol{A}}\left(\beta\right)=\frac{1}{2\beta}\sum_{\ell\in\Z}\sum_{j=0}^2 c_j\left(\frac{i\omega\left(\ell,\beta\right)}{D^{\boldsymbol{A}}_{m_j}+i\omega\left(\ell,\beta\right)}-\frac{i\omega\left(\ell,\beta\right)}{D^{\boldsymbol{A}}_{m_j}-i\omega\left(\ell,\beta\right)}\right.\\
    \left.-\frac{i\omega\left(\ell,\beta\right)}{D^0_{m_j}+i\omega\left(\ell,\beta\right)}+\frac{i\omega\left(\ell,\beta\right)}{D^0_{m_j}-i\omega\left(\ell,\beta\right)}\right)\eqqcolon\frac{1}{2\beta}\sum_{\ell\in\Z}R\left(\omega\left(\ell,\beta\right),\boldsymbol{A}\right)\,,
\end{multline}
on $H^2\left(\R^3,\R^4\right)$, as the latter is the domain of both $\left(D^0_{m_j}\right)^2$ and $\left(D^{\boldsymbol{A}}_{m_j}\right)^2$, for $j=0,1,2$. By iterating six times the second resolvent identity,
\begin{equation*}
    \frac{i\omega\left(\ell,\beta\right)}{D^{ \boldsymbol{A}}_{m_j}\pm i\omega\left(\ell,\beta\right)}-\frac{i\omega\left(\ell,\beta\right)}{D^0_{m_j}\pm i\omega\left(\ell,\beta\right)}=\frac{i\omega\left(\ell,\beta\right)}{D^{\boldsymbol{A}}_{m_j}\pm i\omega\left(\ell,\beta\right)}\left(\boldsymbol{\alpha}\cdot A-V\right)\frac{1}{D^0_{m_j}\pm i\omega\left(\ell,\beta\right)}\,,    
\end{equation*}
one gets
\begin{multline}\label{eq:R-6-terms}
    R\left(\omega\left(\ell,\beta\right),\boldsymbol{A}\right)=\sum_{n=1}^5\left(R_n\left(\omega\left(\ell,\beta\right),\boldsymbol{A}\right)+R_n\left(-\omega\left(\ell,\beta\right),\boldsymbol{A}\right)\right)\\
    +\left(R'_6\left(\omega\left(\ell,\beta\right),\boldsymbol{A}\right)+R'_6\left(-\omega\left(\ell,\beta\right),\boldsymbol{A}\right)\right)\,,
\end{multline}
with
\begin{equation*}\label{eq:R-n}
    R_n\left(\omega\left(\ell,\beta\right),\boldsymbol{A}\right)=\sum_{j=0}^2 c_j\frac{i\omega\left(\ell,\beta\right)}{D^0_{m_j}+i\omega\left(\ell,\beta\right)}\left(\left(\boldsymbol{\alpha}\cdot A-V\right)\frac{1}{D^0_{m_j}+i\omega\left(\ell,\beta\right)}\right)^n
\end{equation*}
and
\begin{equation*}\label{eq:R'-6}
    R'_6\left(\omega\left(\ell,\beta\right),\boldsymbol{A}\right)=\sum_{j=0}^2 c_j\frac{i\omega\left(\ell,\beta\right)}{D^{\boldsymbol{A}}_{m_j}+i\omega\left(\ell,\beta\right)}\left(\left(\boldsymbol{\alpha}\cdot A-V\right)\frac{1}{D^0_{m_j}+i\omega\left(\ell,\beta\right)}\right)^6\,.
\end{equation*}
\begin{remark}\label{stat:KMS-condition}
    Comparing the above argument with the zero-temperature calculations in \cite[Proof~of~Proposition~(3.1)]{GraHaiLewSer-2013-ARMA}, one sees that incorporating a positive temperature simply amounts to performing the following \emph{formal} substitutions:
    \begin{align*}
        \frac{1}{4\pi}\int_\R\diff\omega&\mapsto\frac{1}{2\beta}\sum_{\ell\in\Z}\,,\\
        \omega\in\R&\mapsto\omega\left(\ell,\beta\right)=\frac{\left(2\ell-1\right)\pi}{\beta}\,,\,\ell\in\Z\,,
    \end{align*}
    where $\beta=1/T\in\left(0,+\infty\right)$, and then considering the averaged primitive function with respect to $\beta$ of each term in the expansion. These replacements, already suggested in the study of finite-temperature QED in \cite{Dit-1979-PRD}, make appear the well-known Matsubara frequencies. In the Physics literature, these frequencies usually arise from the KMS condition, which prescribes anti-periodicity with respect to imaginary time for Green's functions of quantum systems at thermal equilibrium. However, in our context, these substitutions stem from a purely mathematical argument relying on the representation formula \eqref{eq:xtanhx}.
\end{remark}

\subsection{Proof of Theorem~\ref{stat:definition-FPV}}
Thanks to Remark~\ref{stat:KMS-condition}, the proof of \cite[Proposition~3.1]{GraHaiLewSer-2013-ARMA} can be adapted to our setting. The estimates rely on Holder's inequality in Schatten spaces and the Kato-Seiler-Simon inequality \cite{SeiSim-1975-CMP}\cite[Theorem~4.1]{Sim-1979-book}, yielding
\begin{multline}\label{eq:TA-trace-estimate}
    \sum_{l\in\Z}\left(\sum_{n=1}^5\norm{\tr_{\C^4}R_n\left(\omega\left(l,\beta\right),\boldsymbol{A}\right)+\tr_{\C^4}R_n\left(-\omega\left(l,\beta\right),\boldsymbol{A}\right)}_{\schatten{1}}\right.\\
    \left.+\norm{\tr_{\C^4}R'_6\left(\omega\left(l,\beta\right),\boldsymbol{A}\right)+\tr_{\C^4}R'_6\left(-\omega\left(l,\beta\right),\boldsymbol{A}\right)}_{\schatten{1}}\right)=\bigO{\beta},
\end{multline}
as $\beta\to0$, where $\bigO{\beta}$ depends on the $4$--potential $\boldsymbol{A}$ through its $L^p$--norms, for $p\in\llbracket 1,5\rrbracket$, as well as the $L^2$--norm of the associated electromagnetic field $\boldsymbol{F}$. Here, $\|\cdot\|_{\schatten{1}}$ denotes the trace-class or $1$--Schatten norm. Since
\begin{align*}
    \mathcal{F}_\mathrm{PV}\left(e\boldsymbol{A},\beta\right)&=\frac{1}{\beta}\int_0^\beta\tr\left(\tr_{\C^4}T_{\boldsymbol{A}}\left(b\right)\right)\diff b\\
    &=\frac{1}{2\beta}\int_0^\beta\tr\left(\sum_{l\in\Z}\tr_{\C^4}R\left(\omega\left(l,b\right),\boldsymbol{A}\right)\right)\frac{\diff b}{b}\,,
\end{align*}
this concludes the proof of item $i)$.\par
Concerning item $ii)$, we remark that the arguments in \cite[Section~3]{GraHaiLewSer-2013-ARMA} also show that the $\schatten{1}$--norms in \eqref{eq:TA-trace-estimate} are actually finite. In particular, we are unable to prove that the $n=1$ and $n=2$ terms are trace-class in the presence of an electrostatic potential without first taking the $\C^4$--trace; however, if $V\equiv0$, it is easy to see that the problematic terms vanish (see \cite[Equations~(3.28)~and~(3.32)]{GraHaiLewSer-2013-ARMA}), so that taking the $\C^4$--trace is no longer necessary to show that they belong to $\schatten{1}$. 

\subsection{Proof of Theorem~\ref{stat:properties-FPV}}\label{sec:main-thm-2}
Combining \eqref{eq:TA-resolvent-difference} with \eqref{eq:R-6-terms}, one can define
\begin{equation*}
    T_{\boldsymbol A}\left(\beta\right)\eqqcolon \sum_{n=1}^5 T_n\left(\boldsymbol A,\beta\right)+T'_6\left(\boldsymbol A,\beta\right)\,.
\end{equation*}
The first step now consists in observing that the odd-order terms vanish and consequently do not contribute to the free energy functional \cite[Lemma~2.1]{BorMor-2026-CAOT}, i.e., for $n=1,3,5$,
\begin{multline*}
    \frac{1}{\beta}\int_0^\beta\tr\left(\tr_{\C^4}\left[T_n\left(\boldsymbol A,b\right)\right]\right)\diff b \\
    =\frac{1}{2\beta}\int_0^\beta\sum_{\ell\in\Z}\tr\left(\tr_{\C^4}\left[R_n\left(\omega\left(\ell,b\right),\boldsymbol A\right)+R_n\left(-\omega\left(\ell,b\right),\boldsymbol A\right)\right]\right)\frac{\diff b}{b}=0\,.
\end{multline*}
This is due to charge-conjugation invariance, which is intrinsic to QED, and corresponds to a well-known result referred to as Furry's theorem in the Physics literature \cite[Section~4.1]{GreRei-2009-book}. Therefore, the PV-regularised free energy functional can be rewritten as
\begin{equation*}\label{eq:PVs}
    \mathcal{F}_\mathrm{PV}\left(\boldsymbol{A},\beta\right)=\mathcal{F}_2\left(\boldsymbol{A},\beta\right)+\mathcal{R}\left(\boldsymbol{A},\beta\right)\,,
\end{equation*}
where
\begin{equation}\label{eq:F2-aux}
    \mathcal{F}_2\left(\boldsymbol{A},\beta\right)\coloneqq\frac{1}{\beta}\int_0^\beta\tr\left(\tr_{\C^4}T_2\left(\boldsymbol A,b\right)\right)\diff b
\end{equation}
and
\begin{equation*}\label{eq:Rs}
    \mathcal{R}\left(\boldsymbol{A},\beta\right)\coloneqq\frac{1}{\beta}\int_0^\beta\tr\left(\tr_{\C^4}T_4\left(\boldsymbol A,b\right)+\tr_{\C^4}T'_6\left(\boldsymbol A,b\right)\right)\diff b\,.
\end{equation*}
The proof of \eqref{eq:R-bound} can be found in \cite[Lemma~2.2]{BorMor-2026-CAOT}. Notice that this bound depends only on the $L^2$--norm of the electromagnetic field $\boldsymbol{F}$ — which is clearly gauge-invariant — thereby improving the bounds of the previous section. This concludes the proof of item $i)$.\par
Item $ii)$ concerns the analysis of $\mathcal{F}_2$ in \eqref{eq:F2-aux}. Focusing on
\begin{equation*}
    T_2\left(\boldsymbol A,\beta\right)=\frac{1}{2\beta}\sum_{\ell\in\Z}\left(R_2\left(\omega\left(\ell,\beta\right),\boldsymbol A\right)+R_2\left(-\omega\left(\ell,\beta\right),\boldsymbol A\right)\right)
\end{equation*}
and applying the $L^2\left(\R^3,\C\right)$-- and the $\C^4$--traces in the right order, one gets
\begin{equation}\label{eq:F2-decomposition-T21-T22-T23}
    \tr\left(\tr_{\C^4}T_2\left(\boldsymbol A,\beta\right) \right)=\mathcal T_{2,1}\left(\boldsymbol A,\beta\right)+\underbrace{\mathcal T_{2,2}\left(A,\beta\right)+\underbrace{\mathcal T^1_{2,2}\left(V,\beta\right)+\mathcal T^2_{2,2}\left(V,\beta\right)}_{\mathcal{T}_{2,2}\left(V,\beta\right)}}_{\mathcal T_{2,2}\left(\boldsymbol A,\beta\right)}+\mathcal T_{2,3}\left(\boldsymbol A,\beta\right)\,,
\end{equation}
where we omit explicit formulae of the terms on the right-hand side to keep the presentation coincise. The interested reader is referred to \cite[Equations~(41)--(47)]{BorMor-2026-CAOT} for full expressions. We emphasise that here the assumption $\operatorname{div} A=0$ is needed to obtain \eqref{eq:F2-decomposition-T21-T22-T23}. The key advantage of splitting the second-order term in this manner is that it allows us to recover both the zero-temperature part, previously derived in \cite{GraHaiLewSer-2013-ARMA}, and the positive-temperature magnetic contribution found in \cite{Mor-2026-AHP}, together with its electrostatic analogue (see \eqref{eq:sum-Txy-M0-MT}). The core idea behind this specific decomposition arises from a technical limitation at finite temperature: following the zero temperature case, one could rewrite the corresponding electrostatic potential contribution $\mathcal T_{2,2}\left(V,\beta\right)$ in the same form as $\mathcal{T}_{2,2}\left(A,\beta\right)$ via an integration by parts with respect to $\omega$. However, at positive temperatures, the discretisation of frequencies prevents us from performing this integration (see Remark~\eqref{stat:KMS-condition}). To circumvent this issue and replicate the argument, we decompose $\mathcal{T}_{2,2}\left(V,\beta\right)$ by adding and subtracting the desired electrostatic term $\mathcal{T}^1_{2,2}\left(V,\beta\right)$. This strategy successfully yields
\begin{multline}\label{eq:sum-Txy-M0-MT}
    \frac{1}{\beta}\int_0^\beta\left(\mathcal{T}_{2,1}\left(\boldsymbol A,b\right)+\mathcal T_{2,2}\left(A,b\right)+\mathcal T^1_{2,2}\left(V,b\right)+\mathcal T_{2,3}\left(\boldsymbol A,b\right)\right)\diff b \\
    =\frac{1}{8\pi}\int_{\R^3}\left(M^0\left(k\right)+M^T\left(k,\beta\right)\right)\left(\vert \widehat B\left(k\right)\vert^2-\vert \widehat E\left(k\right)\vert^2\right) \diff k\,,
\end{multline}
with $M^0$ and $M^T$ respectively given in \eqref{eq:Mzero} and \eqref{eq:MT}, while isolating an additional thermal electrostatic term, introduced in \cite{BorMor-2026-CAOT} and arising from the remainder $\mathcal{T}^2_{2,2}\left(V,\beta\right)$,
\begin{equation*}
    \frac{1}{\beta}\int_0^\beta \mathcal T^2_{2,2}\left(V,b\right)\diff b=\frac{1}{\beta}\int_0^\beta \int_{\R^3}\Gamma\left(k,b\right)\vert \widehat V\left(k\right)\vert^2\diff k\diff b\,,
\end{equation*}
with
\begin{equation}\label{eq:Gamma}
    \Gamma\left(k,\beta\right)=\frac{1}{2\beta\pi^3}\int_{\R^3}\sum^2_{j=0}c_j\sum_{\ell\in\mathbb Z}\frac{\left(3\left(p^2+m_j^2\right)-\omega\left(\ell,\beta\right)^2\right)\omega\left(\ell,\beta\right)^2}{\left(p^2+m^2_j+\omega\left(\ell,\beta\right)^2\right)^2\left(\left(p-k\right)^2+m^2_j+\omega\left(\ell,\beta\right)^2\right)}\diff p\,.
\end{equation}
This concludes the proof of \eqref{eq:F2}. For the sake of brevity, we omit the lengthy and technical calculations that allow us to further decompose $\Gamma$; these can be found in \cite[Subsections~(2.3.1)--(2.3.3)]{BorMor-2026-CAOT}. The analytical properties of $\Gamma$ follow directly from the analysis of each of these resulting terms \cite[Lemma~2.5]{BorMor-2026-CAOT}, concluding the proof of item $ii)$.\par
Item $iii)$ is the content of \cite[Lemma~4.3]{GraHaiLewSer-2013-ARMA} and \cite[Lemma~3.8]{Mor-2026-AHP}.

\subsection{Proof of Theorem~\ref{stat:extension-FPV}}
The bound for the remainder term $\mathcal{R}$ in \eqref{eq:R-bound} allows for its continuous extension to $X$ by density. Therefore, we need to check whether $\mathcal{F}_2$ in \eqref{eq:F2} can be extended from the set of $4$--potentials $\boldsymbol{A}\in L^1\left(\R^3,\R^4\right)\cap\Hdiv$ such that $E=-\nabla V\in L^1\left(\R^3,\R^3\right)$ to $X$ by density. We refer the reader to \cite{Mor-2026-AHP} for the continuous extension of the term involving the Fourier multipliers $M^0\left(k\right)$ and $M^T\left(k,\beta\right)$. For the sake of brevity, we only focus here on the term involving the multiplier $\Gamma\left(k,\beta\right)$. Let $\delta>0$. The following estimate holds:
\begin{align*}
    \frac{1}{\beta}&\int_0^\beta\int_{\R^3}\frac{\Gamma\left(k,b\right)}{\vert k\vert^2}\abs{\widehat{E}\left(k\right)}^2\diff k\diff b\\ 
    &=\frac{1}{\beta}\int_0^\beta\int_{\{\vert k\vert\leq \delta\}}\frac{\Gamma\left(k,b\right)}{\vert k\vert^2}\abs{\widehat{E}\left(k\right)}^2\diff k\diff b +\frac{1}{\beta}\int_0^\beta\int_{\{\vert k\vert\geq \delta\}}\frac{\Gamma\left(k,b\right)}{\vert k\vert^2}\abs{\widehat{E}\left(k\right)}^2\diff k\diff b \\
    &\leq\Vert \widehat E\Vert^2_{L^\infty}\frac{1}{\beta}\int_0^\beta\int_{\{\vert k\vert\leq \delta\}}\frac{\Gamma\left(k,b\right)}{\vert k\vert^2}\diff k\diff b+\Vert \widehat E\Vert^2_{L^2}\left\Vert \frac{\Gamma\left(\cdot,b\right)}{\vert\cdot\vert^2}\right\Vert_{L^\infty(\{\vert k\vert\geq \delta\})}\\
    &\leq C_1\left(\beta\right) \Vert E\Vert_{L^1}+C_2\left(\beta\right)\Vert E\Vert_{L^2}\,, 
\end{align*}
by the Plancherel identity and the continuity of the Fourier transform from $L^1$ to $L^\infty$. Here $C_1(\beta), C_2(\beta)$ are two positive constants, depending on $\beta$. This concludes the proof of Theorem~\ref{stat:extension-FPV}.

\section{Conclusions and Future Perspectives}
In this review, we discussed some recent advances contained in \cite{BorMor-2026-CAOT, Mor-2026-AHP} concerning well-posedness and properties of a free energy functional for the Dirac field interacting with a classical electromagnetic field at thermal equilibrium. The main novelties of the cited works include, on the one hand, a rigorous definition of the free energy both in the purely magnetic and in the electromagnetic case under gauge-invariant assumptions. On the other hand, they involve a mathematical derivation of the thermal contributions appearing in the quadratic term $\mathcal{F}_2$ and representing the linear response of the Dirac field to external electromagnetic fields. Crucially, among these positive-temperature effects, we identify a specific contribution that emerges exclusively in the presence of an external electric field. This term exhibits a singularity which, as discussed in Remark~\ref{rmk:quadratic-term}, we conjecture to be physically linked to a Debye screening effect, in analogy with the result in \cite{HaiLewSei-2008-RMP}. Moreover, the comparison between our positive-temperature calculations and the zero-temperature ones highlights a temperature-induced discretisation of the frequency parameter $\omega$ making appear the Matsubara frequencies usually arising from the KMS condition which prescribes anti-periodicity with respect to imaginary time for Green's functions via the formal replacements in \eqref{eq:T-Feynman-rules}. Instead, our derivation here relies on a purely mathematical argument based on the representation formula \eqref{eq:xtanhx}.\par
For the sake of brevity, we did not discuss the Euler-Heisenberg approximation \cite{HeiEul-1936-ZP} for the free energy functional at positive temperature. Indeed, much like the zero-temperature case, the free energy functional $\mathcal{F}_\mathrm{PV}$ is a complicated non-local functional. One is therefore interested in finding an asymptotic expansion in terms of a superposition of \textit{independent local} functionals, which provides a more tractable description. While a rigorous mathematical derivation of this effective Lagrangian has been first achieved in \cite{GraLewSer-2018-JMPA} in the regime of slowly varying purely magnetic fields at zero temperature, and subsequently extended to incorporate thermal corrections in \cite{Mor-2026-AHP}, a similar result is unfortunately still lacking in the presence of an electrostatic potential, mainly due to the appearance of poles of the functional in the complex plane, which was interpreted by Schwinger as the electron-positron pair production rate \cite{Sch-1951-PR} and explains the instability of the vacuum; see also \cite{DitGie-2000-book}\cite[Paragraph~7.3]{GreRei-2009-book}. Furthermore, we omitted a discussion on the UV convergence of thermal contributions. In the purely magnetic case, these thermal corrections are UV-convergent, suggesting that the results in \cite{Mor-2026-AHP} can be made independent of the PV regularisation parameters by employing the same renormalisation scheme used at zero temperature — a fact indeed proven in the cited work. Conversely, this remains an open problem in the full electromagnetic case.

\section*{Acknowledgements}
The author acknowledges financial support from MUR–Italian Ministry of University and Research and Next Generation EU within PRIN 2022AKRC5P `Interacting Quantum Systems: Topological Phenomena and Effective Theories'. The author also acknowledges financial support from the European Union through the European Research Council’s Starting Grant \textsc{FermiMath}, grant agreement n. 101040991. Views and opinions expressed are those of the authors and do not necessarily reflect those of the European Union or the European Research Council Executive Agency. Neither the European Union nor the granting authority can be held responsible for them. The author has also been partially supported by Gruppo Nazionale per la Fisica Matematica GNFM – INdAM. 

\bibliographystyle{alpha}

\end{document}